\documentstyle[12pt,graphicx,natbib]{article}
\def\sun{\hbox{$\odot$}}

\def\aap{A\&A\,}
\def\aj{AJ}

\begin{document}
\title{
Semi-analytical formulas for the fundamental parameters of
Galactic early B supergiants }
\author
{L. Zaninetti             \\
Dipartimento di Fisica Generale, \\
           Via Pietro Giuria 1   \\
           10125 Torino, Italy
}

\maketitle
\section*{}
The publication of new tables of calibration
of some  fundamental parameters of Galactic
B0-B5 supergiants in the two classes $Ia$ and $Ib$
allow to particularize the eight parameters 
conjecture that model  five fundamental parameters.
The numerical expression for visual magnitude 
,radius , mass , luminosity and surface gravity 
are derived for supergiants in the range of temperature
between   29700 and  15200.
The presence of accurate tables of calibration
allows us  to introduce the efficiency  of the derived formulas.
The average efficiency of the new formulas 
,expressed in percent, 
is 
94  for the  visual magnitude, 
81  for the  mass            , 
96  for the  radius          ,
99  for the logarithm of the luminosity
and 97     for the logarithm of the surface gravity.
\\
keywords                    \\
         {stars: formation       ;
          stars: statistics      ;
          methods: data analysis ;
          techniques: photometric }

\section{Introduction}

A  recent study on the supergiants of spectral type
B0-B5 , see \cite{Searle2008} , reports 
a fine grid of the fundamental parameters 
such as the temperature , luminosity , radius, 
surface gravity and mass.
In this paper we first review the four fundamental 
parameters 
as modeled  
by the eight parameters conjecture 
and we add to the list the surface gravity,
see  Section~\ref{sec_fundamental}.
Section~\ref{secsupergiants} contains the numerical
expression for five fundamental  parameters as well as the
efficiency of such a evaluation.

\section{The fundamental formulas}
\label{sec_fundamental}
 We briefly review the formulas that
characterize the visual magnitude,$M_{V}$ , the  mass
,$\mathcal{M}$,   the radius ,$R$, and the luminosity , $L$,
 of the stars for each MK class as function
 of the  intrinsic , unreddened color index $(B-V)_0$,
see \cite{zaninetti2008c} for details. The first equation models
the visual magnitude , $M_V$
\begin{eqnarray}
M_V=
 - 2.5\,{\it a_{\mathrm {LM}}}- 2.5\,{\it b_{\mathrm {LM}}}\,{\it a_{\mathrm {MT}}}-
\nonumber \\
 2.5\,{\it
b_{\mathrm {LM}}}\,{\it b_{\mathrm {MT}}} \,{\it
log_{10}(\frac{T_{\mathrm{BV}}}{(B-V)_0 - K_{\mathrm{BV}}})}
\nonumber \\
-{\it K_{\mathrm {BC}}}+10\,{\it
log_{10}(\frac{T_{\mathrm{BV}}}{(B-V)_0 - K_{\mathrm{BV}}})}+
\nonumber \\
{\frac {{\it T_{\mathrm {BC}}}}{T_{\mathrm{BV}}}}
 \left [{(B-V)_0 -K_{\mathrm{BV}}} \right] + M_{\mathrm{bol},\sun}
 \quad . \label{eqn_mvbv}
\end{eqnarray}
The second equation connects the mass of the star ,${\mathcal M}$,
with   $(B-V)_0$
\begin{eqnarray}
\label{eqn_mass} \log_{10}
 ( \frac {{\mathcal M}} {\mathcal M}_{\sun})  =
{\it a_{MT}}+\nonumber \\
{\it b_{MT}}\,\ln  \left( {\frac {{\it T_{BV}}}{{\it (B-V)_0}-{\it
K_{BV}} }} \right)  \left( \ln  \left( 10 \right)  \right) ^{-1}
\quad ,
\end  {eqnarray}
where ${\mathcal M}_{\sun}$ is  the sun's  mass. The third
equation regulates the radius $R$, with $(B-V)_0$
\begin{eqnarray}
\label{eqn_radius} \log_{10}
 ( \frac {R}{R{\sun}})  =  \nonumber \\
1/2\,{\it a_{LM}}+1/2\,{\it b_{LM}}\,{\it a_{MT}}+ 2\,{\frac {\ln
\left( {\it T_{\sun}} \right) }{\ln  \left( 10 \right) }}
+\nonumber\\
+1/2\,{\it b_{LM}}\,{\it b_{MT}}\, \ln  \left( {\frac {{\it
T_{BV}}}{{\it {(B-V)_0}}-{\it K_{BV}}}} \right)  \left( \ln  \left(
10 \right)  \right) ^{-1}
\nonumber\\
-2\,\ln  \left( {\frac {{\it T_{BV}} }{{\it {(B-V)_0}}-{\it
K_{BV}}}} \right)  \left( \ln  \left( 10 \right)
 \right) ^{-1}
\quad ,
\end  {eqnarray}
where $R{\sun}$ is the sun's radius. The fourth equation connects
the luminosity of a star $L$ with $(B-V)_0$
\begin{eqnarray}
\label{eqn_luminosity} \log_{10}
 ( \frac {L}{L{\sun}})  =
{\it a_{LM}} +{\it b_{LM}} {\it a_{MT}}    \nonumber \\
+{\it b_{LM}}\left ( {\it b_{MT}}\,\ln  \left( {\frac { {\it
T_{BV}}}{{\it (B-V)_0}-{\it K_{BV}}}} \right)  \frac{1}{\ln  \left(
10
 \right)  }  \right ) ~,
 \end  {eqnarray}
where ${L{\sun}}$ is the sun's luminosity. The eight numerical
parameters that compare above are reported in
Table~\ref{coefficients_formula} as well 
as  the physical or empirical
formula that regulates them.

\begin{table}[h]
\caption{Synoptic Table of the eight coefficients,
 $BC$ is the bolometric correction }
\label{coefficients_formula}
\begin{tabular}{|l|c|}
\hline
Coefficient  &  adopted relationship        \\
                   \hline
$K_{\mathrm{BV}}$~,$T_{\mathrm{BV}}[\mathrm {K}]$               &   $(B-V)_0=  K_{\mathrm{BV}}   + {T_{\mathrm{BV}}}/{T} $ \\
$K_{\mathrm{BC}}$~,$T_{\mathrm {BC}}[\mathrm {K}]$   & $BC  =
-\frac{T_{\mathrm{BC}}}{T} - 10~\log_{10}~T + K_{\mathrm {BC}}$  \\
$a_{\mathrm {LM}}$ ,~$b_{\mathrm {LM}}$              &
$\log_{10}({L}/{L_{\sun}})  = a_{\mathrm {LM}} +b_{\mathrm
{LM}}\log_{10}( {{\mathcal M}}/
 {{\mathcal M}_{\sun}})$ \\
$a_{\mathrm {MT}}$ ,~$b_{\mathrm {MT}}$              & $\log_{10}(
{\mathcal M}/{\mathcal M_{\sun}})  = a_{\mathrm {MT}} +b_{\mathrm
{MT}}\log_{10}({T}
 /{T_{\sun}}) $   \\
\hline
\end{tabular}
\end{table}

A fifth fundamental parameter is the surface 
gravity , $g$ , that is defined as 
\begin{equation}
g = G \frac{M}{r^2} \quad ,
\end{equation}
where $M$ is the mass of the body ,
$r$ its radius 
and  $G$ is the
Newtonian gravitational constant 
which has value  $G=6.6742 \times 10^{-11}\frac { m^3} {kg
s^2}$, ~\cite{CODATA2005}.
On adopting $R_{sun}=6.95508 \cdot 10^8~m$
and  $M_{sun}=1.989 \cdot10^{30}~kg$ , see \cite{cox},
we obtain
the following expression 
for the logarithm of the surface gravity
\begin{eqnarray}
log(g[cgs])=- 10.60+ 0.4342\times
\nonumber  \\ 
\times \ln  \left ( {{\rm e}^{ 2.302\,{\it 
a_{MT}}- 2.302\,{\it a_{LM}}- 2.302\,{\it b_{LM}}\,{\it a_{MT}}}}
 \left( {\frac {{\it T_{BV}}}{{\it _{BV}}- {\it K_{BV}}}}\right ) ^{{\it 
b_{MT}}- {\it b_{LM}}\,{\it b_{MT}}+ 4} \right) 
\quad .
\label{surfaceg}
\end{eqnarray}

\section{Application to the supergiants}
\label{secsupergiants}
The eight parameters conjecture may represents an acceptable fit
of five fundamental parameters of the stars once the calibration
data are available in the considered MK class, in our case
B0-B5~supergiants. The Table 5 in \cite{Searle2008} provides the
calibration of $\log_{10} ( \frac {L}{L{\sun}}) $, $\log_{10}
 ( \frac {R}{R{\sun}})$ , $\frac {{\mathcal M}} {\mathcal
 M}_{\sun}$ and $\log_{10}(g [cm/s^2])$ as
 function of the temperature $T$ in the range
 $29700 <T <15200$.
Table~\ref{coefficients_supergiants} reports the eight
 parameters
as well as the source where the calibrated data resides.

\begin{table}[h]
\caption{Table of the adopted coefficients for B0-B5~supergiants
.} \label{coefficients_supergiants}
\begin{tabular}{|l|c|c|c|}
\hline
Coefficient  &  Ia   & Ib &  source~of~data        \\
                   \hline
$K_{\mathrm{BV}}$              & -0.3961 &  -0.3961 &         Table~ 15.7~ in~ Cox ~(2000)    \\
$T_{\mathrm{BV}}[\mathrm {K}]$ &  4011.7 & 4011.7   & Table~ 15.7~ in~ Cox ~(2000)  \\
$K_{\mathrm{BC}}$              & 42.87   &  42.87   & Table~ 15.7~ in~ Cox ~(2000)    \\
$T_{\mathrm {BC}}[\mathrm {K}]$& 31573.8 & 31573.8  & Table~ 15.7~ in~ Cox ~(2000)  \\
$a_{\mathrm {LM}}$             &  4.667  & 4.092   & Table~5~in~  Searle~et~al.~(2008) \\
$b_{\mathrm {LM}}$             & 0.6050  &  0.92054& Table~5~in~  Searle~et~al.~(2008)    \\
$a_{\mathrm {MT}}$             & -3.713  & -6.0246  & Table~5~in~  Searle~et~al.~(2008)   \\
$b_{\mathrm {MT}}$             &  1.1674 &  1.7213 & Table~5~in~  Searle~et~al.~(2008)     \\
\hline
\end{tabular}
\end{table}
The fundamental  parameters of the stars are parametrized 
according to  the MK class to which they belong and 
the  intrinsic , unreddened color index  $(B-V)_0$ or the temperature 
as derived , for example , from spectroscopic arguments.
The conversion between temperature and $(B-V)_0$ is obtained
trough the following two  formulas
\begin{eqnarray}
(B-V)_0 =  - 0.3961+ \frac{4011.7}{T}  \\
 15200~K <T< 30000~K      \quad ,           \nonumber
\end{eqnarray}
\begin{eqnarray}
T  =  \frac {
4011.7 }
{(B-V)_0  + 0.3961 } 
\\
-0.25   <(B-V)_0<  -0.14 \quad . \nonumber
\nonumber
\end{eqnarray}

\begin{figure}
\begin{center}
\includegraphics[width=10cm]{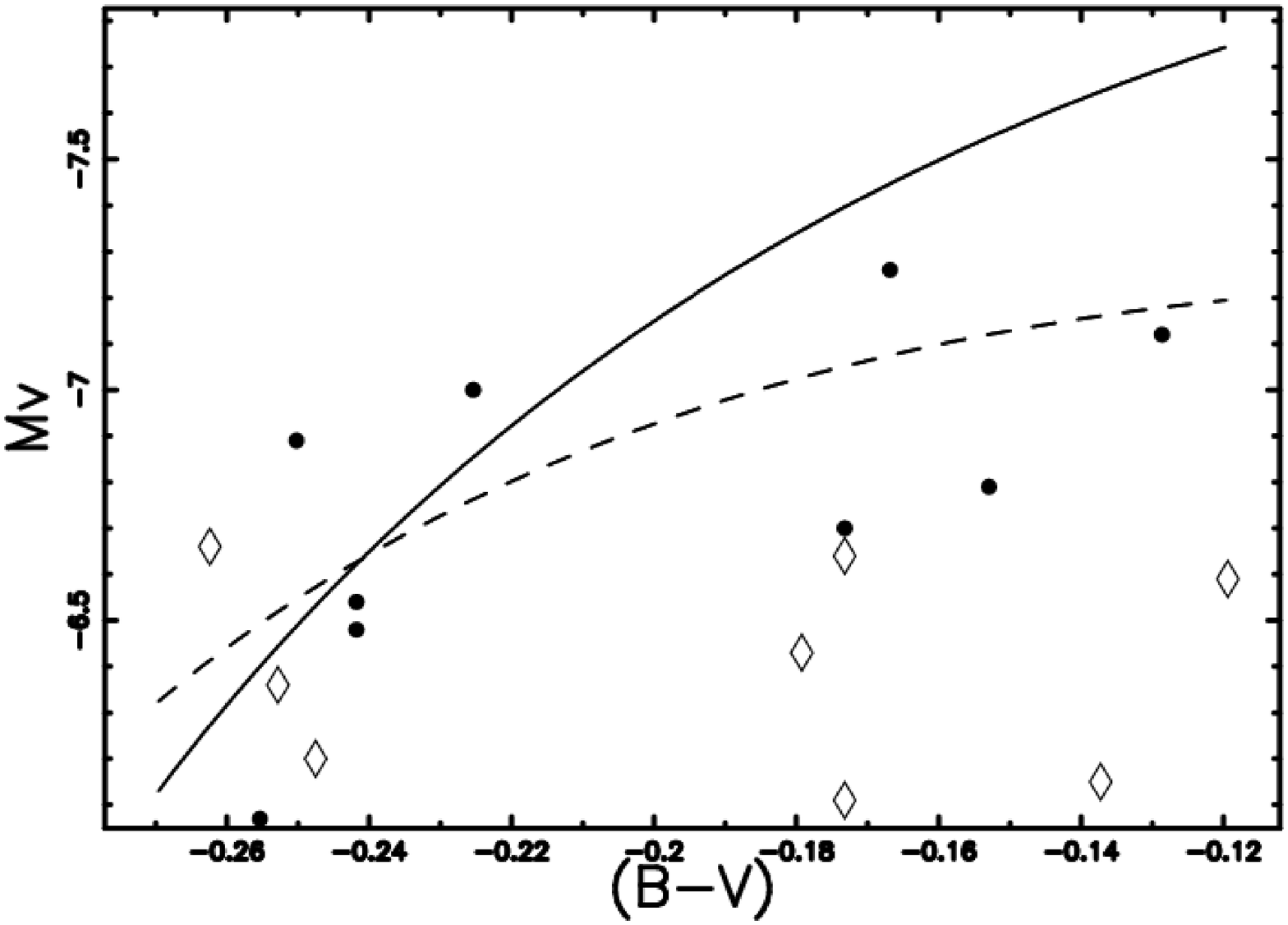}
\end{center}
\caption{Theoretical visual magnitude ,$M_{\mathrm V}$ ,
 against $(B-V)_0$
for Galactic early B supergiants   : Ia (full line)
and Ib (dotted line).
The observed values of magnitude as extracted from Table 3 of 
Searle~2008 are also reported :
Ia (full circle ) and Ib (empty diamond). 
  }
\label{f01}
\end{figure}

From a numerical point of view  
the visual magnitude , $M_{\mathrm V}$ , is 
\begin{eqnarray}
M_V=
- 41.07+ 3.576\,\ln  \left(  4012.0\, \left( {\it (B-V)_0}+ 0.3961 \right) 
^{-1} \right) + 7.870\,{\it (B-V)_0}
\\
supergiants~Ia~ when~0.25   <(B-V)_0<  -0.14 \quad , 
\nonumber
\label{mvia}
\end{eqnarray}
and
\begin{eqnarray}
M_V=
- 31.38+ 2.623\,\ln  \left(  4012.0\, \left( {\it (B-V)_0}+ 0.3961 \right) 
^{-1} \right) + 7.870\,{\it (B-V)_0}
\\
supergiants~Ib~ when~0.25   <(B-V)_0<  -0.14 \quad . 
\nonumber
\label{mvib}
\end{eqnarray}
Figure ~\ref{f01} reports the theoretical 
visual magnitude ,
$M_V$ , for the two classes here considered as well as 
the observational points as extracted from Table~3 of 
\cite{Searle2008}.

\begin{figure}
\begin{center}
\includegraphics[width=10cm]{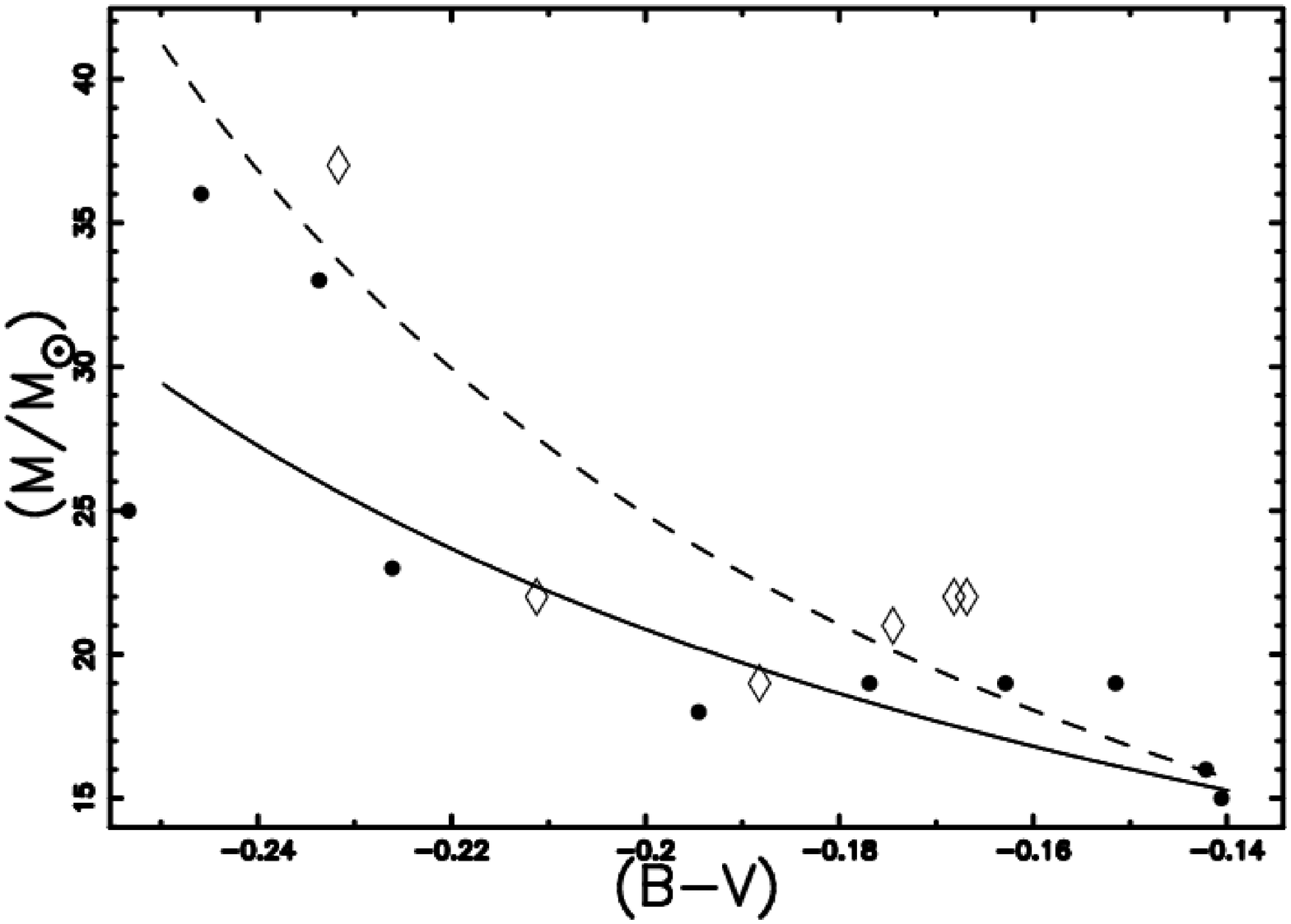}
\end{center}
\caption{Theoretical mass , 
$\frac{{\mathcal{M}}} {{\mathcal{M}_{\sun}}}$ ,
 against $(B-V)_0$
for Galactic early B supergiants   : Ia (full line)
and Ib (dotted line).
The observed values of  mass as extracted from Table 4 of 
Searle~2008 are also reported :
Ia (full circle ) and Ib (empty diamond). 
  }
\label{f02}
\end{figure}

The mass $\frac{{\mathcal{M}}} {{\mathcal{M}_{\sun}}}$
has  expression
\begin{eqnarray}
\frac{{\mathcal{M}}} {{\mathcal{M}_{\sun}}} =
{ 10.0}^{- 3.713+ 0.5070\,\ln  \left(  4012.0\, \left( {\it (B-V)_0}+
 0.3961 \right) ^{-1} \right) }
\\
supergiants~Ia~ when~0.25   <(B-V)_0<  -0.14 \quad ,
\nonumber
\end{eqnarray}
and
\begin{eqnarray}
\frac{{\mathcal{M}}} {{\mathcal{M}_{\sun}}} =
{ 10.0}^{- 6.025+ 0.7476\,\ln  \left(  4012.0\, \left( {\it (B-V)_0}+
 0.3961 \right) ^{-1} \right) }
\\
supergiants~Ib~ when~0.25   <(B-V)_0<  -0.14 \quad .
\nonumber
\end{eqnarray}
Figure ~\ref{f02} reports the theoretical 
mass ,
$\frac{{\mathcal{M}}} {{\mathcal{M}_{\sun}}}$,
 for the two classes here considered as well as  
the observational points as extracted from Table~4 of 
\cite{Searle2008}.

\begin{figure}
\begin{center}
\includegraphics[width=10cm]{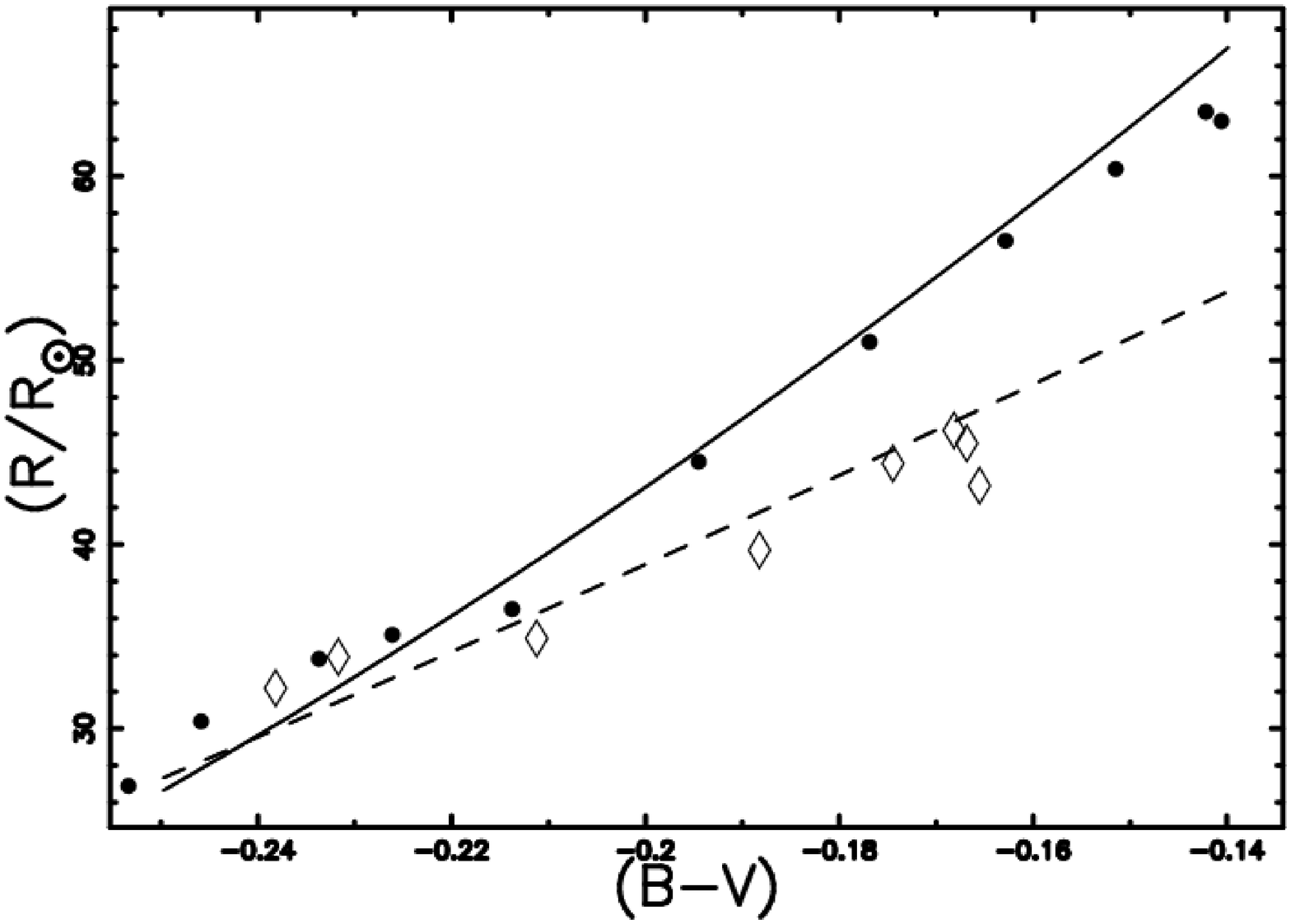}
\end{center}
\caption{Theoretical 
radius , $\frac{{{R}}} {{{R}_{\sun}}}$ , against $(B-V)_0$
for Galactic early B supergiants   : Ia (full line)
and Ib (dotted line).
The observed value of radius   as extracted from Table 4 of 
Searle~2008 are also reported :
Ia (full circle ) and Ib (empty diamond). 
  }
\label{f03}
\end{figure}

The radius , $\frac{{{R}}} {{{R}_{\sun}}}$,
has expression 
\begin{eqnarray}
\frac{{{R}}} {{{R}_{\sun}}}=
{ 10.0}^{ 8.733- 0.7151\,\ln  \left(  4012.0\, \left( {\it (B-V)_0}+ 0.3961
 \right) ^{-1} \right) }
\\
supergiants~Ia~ when~0.25   <(B-V)_0<  -0.14 \quad ,
\nonumber
\end{eqnarray}
and
\begin{eqnarray}
\frac{{{R}}} {{{R}_{\sun}}}=
{ 10.0}^{ 6.795- 0.5245\,\ln  \left(  4012.0\, \left( {\it (B-V)_0}+ 0.3961
 \right) ^{-1} \right) }
\\
supergiants~Ib~ when~0.25   <(B-V)_0<  -0.14 \quad .
\nonumber
\end{eqnarray}
Figure ~\ref{f03} reports the theoretical 
radius ,
$\frac{{{R}}} {{{R}_{\sun}}}$,
 for the two classes here considered as well as 
the observational points as extracted from Table~4 of 
\cite{Searle2008}.

\begin{figure}
\begin{center}
\includegraphics[width=10cm]{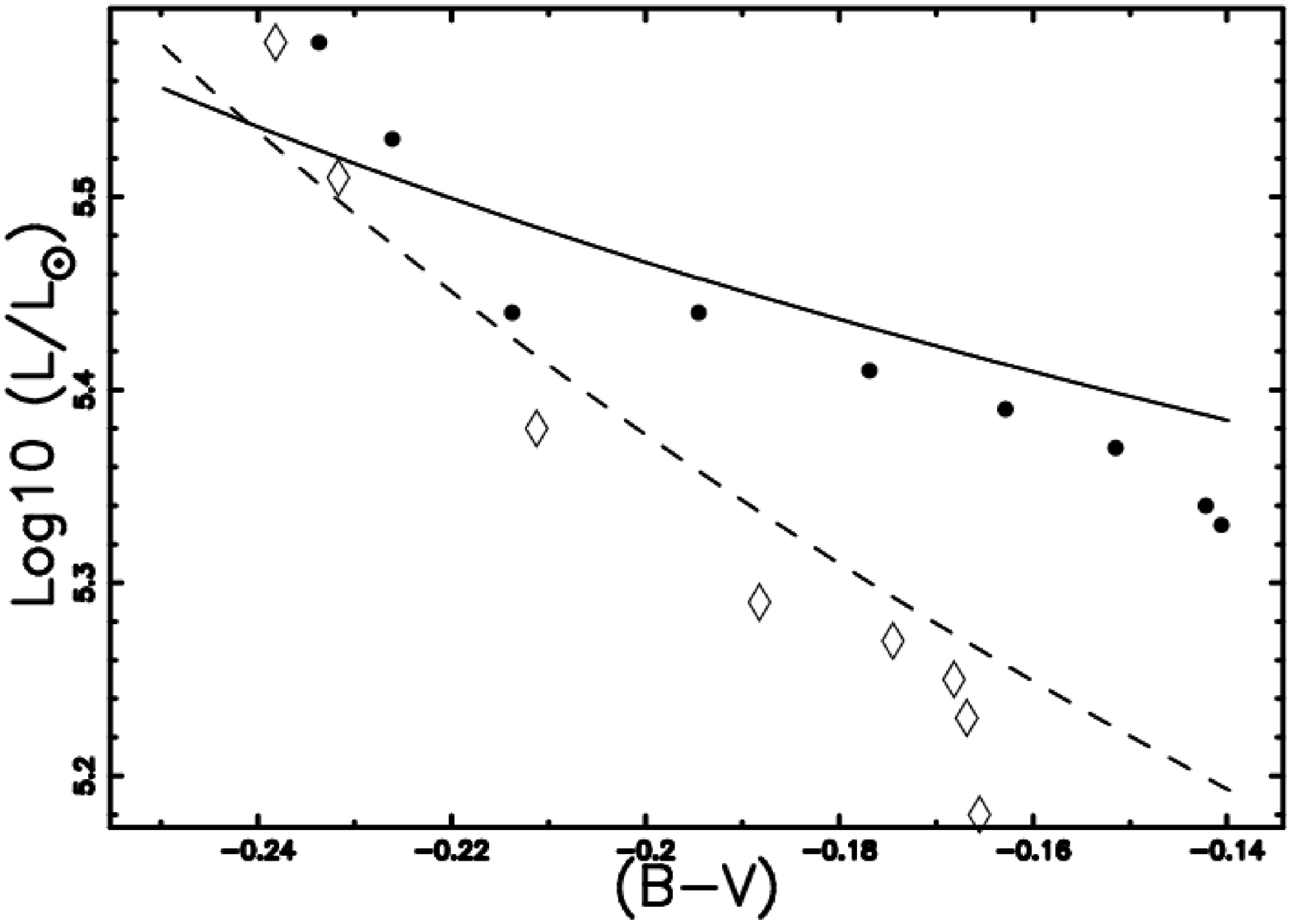}
\end{center}
\caption{Theoretical logarithm of 
the luminosity 
 , $\log_{10}
 ( \frac {L}{L{\sun}}) $ , against $(B-V)_0$
for Galactic early B supergiants   : Ia (full line)
and Ib (dotted line).
The observed values of luminosity as extracted from Table 4 of 
Searle~2008 are also reported :
Ia (full circle ) and Ib (empty diamond). 
  }
\label{f04}
\end{figure}

The logarithm of the luminosity ,
$\log_{10} ( \frac {L}{L{\sun}}) $,
is 
\begin{eqnarray}
\log_{10} ( \frac {L}{L{\sun}}) =
2.421+ 0.3068\,\ln  \left(  4012.0\, \left( {\it (B-V)_0}+ 0.3961 \right) 
^{-1} \right) 
\\
supergiants~Ia~ when~0.25   <(B-V)_0<  -0.14 \quad ,
\nonumber
\end{eqnarray}
and
\begin{eqnarray}
\log_{10} ( \frac {L}{L{\sun}}) =
- 1.454+ 0.6882\,\ln  \left(  4012.0\, \left( {\it (B-V)_0}+ 0.3961
 \right) ^{-1} \right) 
\\
supergiants~Ib~ when~0.25   <(B-V)_0<  -0.14 \quad .
\nonumber
\end{eqnarray}
Figure ~\ref{f04} reports the theoretical 
logarithm of the luminosity ,
$\log_{10} ( \frac {L}{L{\sun}}) $,
 for the two classes here considered as well as 
the observational points as extracted from Table~4 of 
\cite{Searle2008}.

\begin{figure}
\begin{center}
\includegraphics[width=10cm]{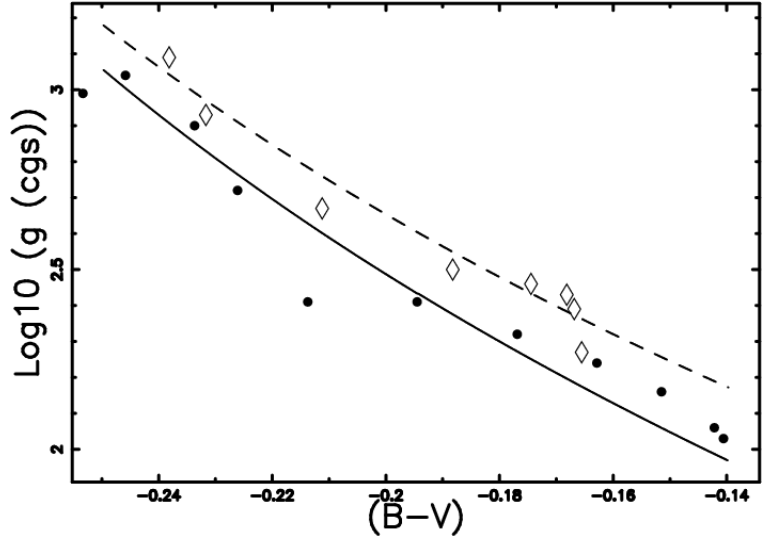}
\end{center}
\caption{Theoretical logarithm of 
the surface gravity
 , $\log_{10} ( g [cgs]) $ , against $(B-V)_0$
for Galactic early B supergiants   : Ia (full line)
and Ib (dotted line).
The observed values of surface gravity  as extracted from Table 4 of 
Searle~2008 are also reported :
Ia (full circle ) and Ib (empty diamond). 
  }
\label{f05}
\end{figure}

The logarithm of the surface gravity  ,
$ \log_{10} ( g [cgs]) $ , is 
\begin{eqnarray}
\log_{10} ( g [cgs]) =   \nonumber \\
- 10.61+ 0.4343\,\ln  \left(  3.118\, \left(  \left( {\it (B-V)_0}+ 0.3961
 \right) ^{-1} \right) ^{ 1.167} \right) 
   \nonumber \\
- 0.8686\,\ln  \left( 
 0.00001891\, \left(  \left( {\it (B-V)_0}+ 0.3961 \right) ^{-1} \right) ^{
- 1.646} \right) 
\\
supergiants~Ia~ when~0.25   <(B-V)_0<  -0.14 \quad ,
\nonumber
\label{surfacegia}
\end{eqnarray}
and
\begin{eqnarray}
\log_{10} ( g [cgs])=   \nonumber \\
- 10.61+ 0.4343\,\ln  \left(  1.506\, \left(  \left( {\it (B-V)_0}+ 0.3961
 \right) ^{-1} \right) ^{ 1.721} \right)
                         \nonumber\\
 - 0.8686\,\ln  \left( 
 0.000008342\, \left(  \left( {\it (B-V)_0}+ 0.3961 \right) ^{-1} \right) ^
{- 1.2077} \right) 
\\
supergiants~Ib~ when~0.25   <(B-V)_0<  -0.14 \quad .
\nonumber
\label{surfacegib}
\end{eqnarray}
Figure ~\ref{f05} reports the theoretical 
logarithm of the  surface gravity 
for the two classes here considered as well as  
the observational points as extracted from Table~4 of 
\cite{Searle2008}.
From a practical point  of view,  
$\epsilon$ ,
the percentage  of
reliability  of our results can also be  introduced,
\begin{equation}
\epsilon  =(1- \frac{\vert( F_{obs}- F_{num}) \vert}
{F_{obs}}) \cdot 100
\,,
\label{efficiency}
\end{equation}
where $F_{obs}$ is one of  the 
fundamental   parameter  as given
by the astronomical observations  ,
and  
$F_{num}$   the  analogous  
fundamental   parameter  as given
by our numerical relationships. 
The minimum  , averaged and maximum efficiency 
in  reproducing
the observed fundamental  parameters  
( 
respectively 
$\epsilon_{min}$    , 
$\overline{\epsilon}$ and  
$\epsilon_{max}$ 
)
as given by formula~(\ref{efficiency} )
are reported in Table~\ref{tab:fundamental}.

\begin{table}[ht!]
\caption {Efficiency in deriving the fundamental 
parameters  for B0-B5~supergiants}
\label{tab:fundamental}
\begin{center}
\begin{tabular}{|c|c|c|c|c|c|c|}
\hline
~ & \multicolumn{3}{|c|} {Ia} & \multicolumn{3}{|c|} {Ib}  
\\
\hline
Fundamental Parameter                     
                      & $ \epsilon_{min}~(\%)       $
                      & $ \overline{\epsilon}~(\%)  $
                      & $\epsilon_{max}~(\%)       $
                      & $\epsilon_{min}~(\%)       $
                      & $\overline{\epsilon}~(\%)  $
                      & $\epsilon_{max}~(\%)       $
 \\
\hline   
Visual~magnitude & 88.8  & 94.5 & 98.7 & 83.5 & 91.4 & 97.5 \\
Mass & 10.5  &81.2 & 97.7 & 55.2 & 81.4 & 95.9 \\
Radius & 91.5  & 95.9 & 98.5 & 90.4 & 94.8 & 98.9 \\
Logarithm~Luminosity & 98.7  &99.2 & 99.6 & 98.3 & 99.1 & 99.6 \\
Logarithm~surface~gravity & 90.9  &96.8 & 98.9 & 95.9 & 98.2 & 99.5 \\
\hline 
\end{tabular}
\end{center}
\end{table}

\section{Conclusions}
The eight parameters conjecture which  allows 
to model five fundamental parameters of the stars 
is connected with the availability of calibration 
data on mass and  luminosity as a function of 
the temperature or the intrinsic , unreddened color index  
$(B-V)_0$.
Table 3 and 4  in \cite{Searle2008} offer  both
the splitting of early supergiants in two classes 
and a good coverage 
of the fundamental  parameters in the
range 
$15200~K <T< 30000~K$ .
The five fundamental  parameters here analyzed 
can be parametrized as function of $(B-V)_0$ 
and  Table~\ref{tab:fundamental} reports 
the efficiency of the model here implemented.
As an example the two relationships (\ref{mvia})
and (\ref{mvib}) that  regulates the visual
magnitude have an averaged accuracy of 
$0.34 ~mag$ in Ia  
and $0.5~mag$ in  Ib.
This paper also parametrizes
the logarithm of the surface gravity 
 , $\log_{10} ( g [cgs]) $ , as a function $(B-V)_0$,
see formulas~(\ref{surfaceg}),
             (\ref{surfacegia}),
             (\ref{surfacegib}) and  
             Figure~\ref{f05}.
This parametrization allows to casts doubt 
of the fits of $T$ that contain   
$\log_{10} ( g [cgs]) $ as a parameter 
, see for example \cite{Sekiguchi2000,Kovtyukh2008} ,
because the reverse is true:
$\log_{10} ( g [cgs]) $ is a function of $T$ or its 
observational counterpart  $(B-V)_0$.

{\bf Acknowledgements}
I thank Lorenzo Ducci who
drew my attention to
Table 5 in \cite{Searle2008}.


\begin{thebibliography}{6}
\expandafter\ifx\csname natexlab\endcsname\relax\def\natexlab#1{#1}\fi

\bibitem[{{Cox}(2000)}]{cox}
{Cox}, A.~N. {: 2000}, {Allen's astrophysical quantities} (New York: Springer)

\bibitem[{{Kovtyukh} {et~al.}(2008){Kovtyukh}, {Soubiran}, {Belik},
  {Yasinskaya}, {Chekhonadskikh}, \& {Malyuto}}]{Kovtyukh2008}
{Kovtyukh}, V.~V., {Soubiran}, C., {Belik}, S.~I., {et~al.} {: 2008},
  Kinematics and Physics of Celestial Bodies, {\bf 24}, 171

\bibitem[{{Mohr} \& {Taylor}(2005)}]{CODATA2005}
{Mohr}, P.~J. \& {Taylor}, B.~N. {: 2005}, Reviews of Modern Physics, {\bf 77},
  1

\bibitem[{{Searle} {et~al.}(2008){Searle}, {Prinja}, {Massa}, \&
  {Ryans}}]{Searle2008}
{Searle}, S.~C., {Prinja}, R.~K., {Massa}, D., \& {Ryans}, R. {: 2008}, \aap,
  {\bf 481}, 777

\bibitem[{{Sekiguchi} \& {Fukugita}(2000)}]{Sekiguchi2000}
{Sekiguchi}, M. \& {Fukugita}, M. {: 2000}, \aj, {\bf 120}, 1072

\bibitem[{{Zaninetti}(2008)}]{zaninetti2008c}
{Zaninetti}, L. {: 2008}, Serbian Astronomical Journal, {\bf 177}, 73

\end{thebibliography}

\end{document}